\documentclass[prb,reprint, amsmath,amssymb,
 aps,superscriptaddress,floatfix]{revtex4-2}

\usepackage{amsmath,amsthm,amsfonts,amssymb,mathrsfs} 
\usepackage[colorlinks=true,linkcolor=blue,urlcolor=blue,citecolor=blue]{hyperref}

\usepackage{graphicx}
\graphicspath{{./images/}}

\usepackage{xcolor}
\usepackage{tikz}
\usepackage{comment}

\newcommand{\bra}[1]{\left\langle#1\right|}
\newcommand{\ket}[1]{\left|#1\right\rangle}
\newcommand{\abs}[1]{\left|#1\right|}
\newcommand{\spacecomma}{\hspace{3mm},\hspace{3mm}}
\DeclareMathOperator{\Tr}{Tr}

\begin{document}

\title{Information lattice approach to the metal-insulator transition}
\author{William Skoglund}
\affiliation{NanoLund and Division of Mathematical Physics, Department of Physics, Lund University, Lund, Sweden}

\author{Elton Giacomelli}
\affiliation{NanoLund and Division of Mathematical Physics, Department of Physics, Lund 
University, Lund, Sweden}

\author{Yiqi Yang}
\affiliation{NanoLund and Division of Mathematical Physics, Department of Physics, Lund University, Lund, Sweden}

\author{Jens H. Bardarson}
\affiliation{Department of Physics, KTH Royal Institute of Technology, 106 91, Stockholm, Sweden}

\author{Erik van Loon}
\affiliation{NanoLund and Division of Mathematical Physics, Department of Physics, Lund University, Lund, Sweden}

\begin{abstract}
Correlation functions and correlation lengths are frequently used to describe phase transitions in quantum systems, but they require an explicit choice of observables. The recently introduced information lattice instead provides an observable-independent way to identify where and at which scale information is contained in quantum lattice models. Here, we use it to study the difference between the metallic and insulating regime of one-dimensional noninteracting tight-binding chains. We find that the information per scale follows a power law in metals at low temperature and that Friedel-like oscillations are visible in the information lattice. At high temperature or in insulators at low temperature, the information per scale decays exponentially. Thus, the information lattice is a useful tool for analyzing the metal-insulator transition.
\end{abstract}

\maketitle

%\EvL{Refs (check if we should include them):
%Metal-insulator transition from the density-matrix perspective, Friedel oscillations~\cite{Vonsovskii_1989}.
%Information approach for the Hubbard model \cite{bellomia2025localclassicalcorrelationsphysical}.
%Walter Kohn paper on many-particle wavefunction of the insulating state \cite{PhysRev.133.A171}.
%Local Localization marker\cite{Marrazzo2019}, in terms of wavefunction.
%(references provided by Gabriele Bellomia)
%}

\section{Introduction}

Quantum mechanics is essential for understanding the electronic properties of solids, since the delocalization of electrons is behind phenomena such as band structures, chemical bonding and metallicity~\cite{hofmann2022solid}. Still, a description using electronic quasiparticles and relatively little genuine many-particle quantum mechanics is sufficient to understand the properties of many standard materials. In fact, the widely used density functional theory~\cite{DFToverview} has been shown to have predictive power for wide classes of materials, even though it only works with the electron density~\cite{HohenbergKohn64} instead of the full many-electron wavefunction.

Moving beyond the quasiparticle picture, strong electronic correlations in quantum materials~\cite{basov2017towards} result in unconventional phases with unique properties, but also require methodology that goes beyond density functional theory. Due to the difficulty of the many-electron problem, an exact solution is often out of reach and approximations have to be made~\cite{LeBlanc15}. The expected spatial structure of the electronic properties provides a way to make approximations. One possibility is to directly simulate finite size clusters to approximate the thermodynamic limit. An alternative to reduce the finite size effects is to couple the cluster to a bath, as in cluster extensions of dynamical mean-field theory~\cite{Georges96,Maier05}. In the same spirit, quantum embedding approaches~\cite{sun2016quantum} in quantum chemistry make a spatial separation into regions with strong and weak quantum effects and corresponding high and low level computations. Another important example of scale-based approximations occurs in the study of surfaces and interfaces, where electronic structure calculations of finite-size slabs~\cite{sun2013efficient} are used to represent a semi-infinite bulk. 

\begin{figure}
    \centering
    \includegraphics{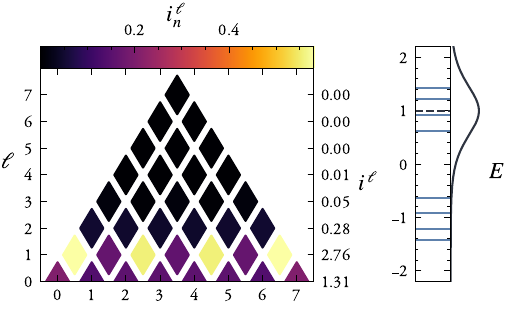}
    \vspace{-5mm}
\definecolor{rust}{HTML}{B7410E}
\definecolor{darkblue}{HTML}{2E3440}
\begin{tikzpicture}[x=3.375in, y=1cm, trim left=0in,
    siteA/.style={circle, minimum size=3mm, draw=black, fill=white},
    siteB/.style={circle, minimum size=3mm, draw=black, fill=black!50},
    hoppingInter/.style={thin},
    hoppingIntra/.style={-, very thick}
]

% Lattice Parameters
\def\plotWidth{0.54} %Based on matplotlib gs
\def\nCells{3} % Actually # cells - 1
\pgfmathsetmacro{\xOffset}{-0.49 + \plotWidth/2}
\pgfmathsetmacro{\baseSpacing}{\plotWidth/(\nCells+1)}
\pgfmathsetmacro{\siteSpacing}{\baseSpacing/2}
\def\subLatticeOne{0}
\def\subLatticeTwo{0}
\def\labelpadhop{1.8mm}
\def\labelpadV{1.5mm}

\foreach \n in {0,...,\nCells} {
    \pgfmathsetmacro{\xpos}{\xOffset + \n * \baseSpacing}
    \node[siteA] (A\n) at (\xpos - \siteSpacing/2, \subLatticeOne) {};
    \node[siteB] (B\n) at (\xpos + \siteSpacing/2, \subLatticeTwo) {};

    \ifnum \n>1
        \node[below=\labelpadV] at (A\n) {V};
        \node[below=\labelpadV] at (B\n) {-V};
    \fi
}

\foreach \n in {0,...,\nCells} {
    \pgfmathtruncatemacro{\next}{\n+1}
    \ifnum \n<2
        \draw[hoppingIntra] (A\n) to[] node[below=\labelpadhop] {$t_1$} (B\n);
    \else
        \draw[hoppingIntra] (A\n) to[] node[] {} (B\n);
    \fi 
    \ifnum \n<\nCells
        \ifnum \n<1
            \draw[hoppingInter] (B\n) to node[below=\labelpadhop] {$t_2$} (A\next);
        \else
            \draw[hoppingInter] (B\n) to node[] {} (A\next);
        \fi
    \fi
}
\end{tikzpicture}
    \caption{Sketch of Rice-Mele Hamiltonian of length $N=8$ (bottom left), with weak and strong bonds $t_1$, $t_2$ and alternating on-site potential $V$. The corresponding information lattice can be visualized as a pyramid (top left), here for $t_1=1$, $t_2=0.5$, $V=0$, at $k_B T=1/3$ and $\mu=1$. It shows that most information is contained at the level of the dimers ($\ell=1$ and strong bonds). The row sum $i^\ell$ of the information is shown on the right axis. The electronic spectrum (panel right) for this short chain is discrete; the formation of continuous bands happens in the limit of large $N$. The curve next to the spectrum shows $n_\text{FD}(1-n_\text{FD})$ to illustrate which energy levels are partially occupied at this temperature and chemical potential $\mu$.
    }
     \label{fig:small}
\end{figure}
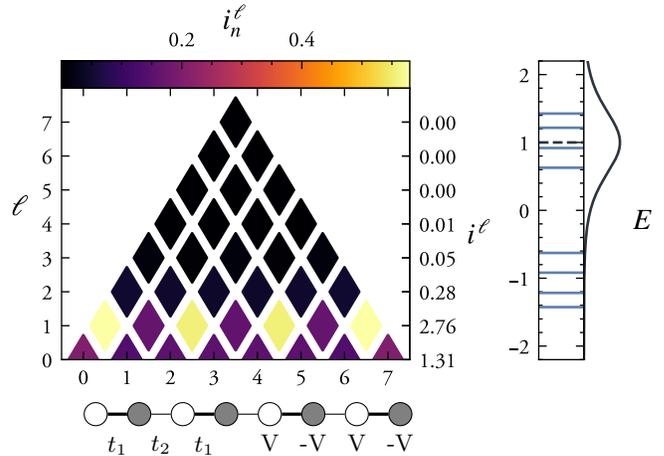

The prevalence of this kind of scale-based approximations, where brute force convergence with respect to system size is challenging~\cite{xu2024coexistence}, means that it is useful to quantify the relevant length scales in electronic systems: when is the cluster, slab, or region large enough? A traditional way to estimate the required length scale is based on the correlation functions of specific operators and the associated correlation length $\xi$. Convergence requires that the cluster size exceeds the correlation length. A disadvantage of using correlation functions is that it requires an explicit choice of operators. Especially in complex quantum systems, the proper choice of operators might not be \emph{a priori} obvious. The information lattice~\cite{Kvorning2022, Artiaco_2024, Flor2025} is a recently proposed method to describe the location and length scale of information in quantum systems. Since it is based on information instead of observables, it identifies the relevant length scales in an unbiased way. The information lattice has previously been applied to various state characterization and analysis of dynamics~\cite{Artiaco25, Artiaco2025b, Harkins2025, Bauer25, Bilinskaya2025, nava2026information}.
The information lattice is part of a wider effort of characterizing electronic properties using information concepts~\cite{zeng2019quantum,LAFLORENCIE20161, ors2019tensor, laurell2025witnessing}. 

Here we use the information lattice for one-dimensional noninteracting electronic systems, in the form of tight-binding chains. Whether the system behaves as a metal or an insulator depends on the position of the chemical potential with respect to the band gap. The metal and insulator are associated with delocalized and localized electrons, respectively, which leads to drastic changes in the information lattice. At low temperature, we show that the information as a function of scale shows power law decay in the metal and exponential decay in the band insulator. At higher temperature, comparable to the size of the gap, exponential decay is found regardless of the chemical potential. Apart from the length scale, the position within the finite chain also reflects the physics of the system. The metal has Friedel-like oscillations appearing in the information lattice, while the presence or absence of edge modes is visible for insulators. These results show that the information lattice provides a useful view on the metal-insulator phase transition and the effect of temperature in electronic systems.

\section{Methods}

\subsection{Model}

We consider the one-dimensional spinless noninteracting fermionic Rice-Mele model, with Hamiltonian
\begin{align}
 H = \sum^{N-2}_{\substack{i=0,\\ i \text{ even}}} \bigg[ &-t_1 c^\dagger_i c^{\phantom{\dagger}}_{i+1}- t_1 c^\dagger_{i+1} c^{\phantom{\dagger}}_{i} -t_2 c^\dagger_{i+1} c^{\phantom{\dagger}}_{i+2}-t_2 c^\dagger_{i+2} c^{\phantom{\dagger}}_{i+1} \notag \\
  &+ V c^\dagger_i c^{\phantom{\dagger}}_i - V c^\dagger_{i+1} c^{\phantom{\dagger}}_{i+1} \bigg].
\end{align}
Here, $N$ is the total number of sites. 
We allow $N$ to be even or odd. 
We use open boundary conditions so terms in the Hamiltonian with creation or annihilation operators on or beyond site $N$ are excluded. 
The hopping parameters $t_1$, $t_2$ and the on-site potential $V$ are all assumed to be real. 
The spectrum is discrete and symmetric around $E=0$; an example is shown in the right panel of Fig.~\ref{fig:small}. 
In the thermodynamic limit, bands form and the model has a gap of magnitude $2\sqrt{V^2+(t_1-t_2)^2}$ (see App.~\ref{app:periodic}), which remains a good estimate for large finite $N$.

For $V=0$, this model reduces to the Su-Schrieffer-Heeger (SSH) model~\cite{SSHmodel,SSHmodel2}, while it is a simple staggered potential for $t_1=t_2$, which are textbook examples~\cite{asboth2016short} of models with topological and trivial band gaps, respectively. The examples shown in this work are either purely SSH ($V=0$) or purely staggered potential ($t_1=t_2$). The electron density $\langle c^\dagger_i c^{\phantom{\dagger}}_i \rangle$ is uniform in the SSH model in the thermodynamic limit, while it alternates in the staggered model: for $V>0$, the density is larger on even sites with local potential $-V$ and smaller on odd sites with local potential $+V$. 
The bond kinetic energies $\langle c^\dagger_i c^{\phantom{\dagger}}_{i+1}+c^\dagger_{i+1} c^{\phantom{\dagger}}_{i}\rangle$ are the same for all bonds in the staggered model but not in the SSH model where dimerization takes place.
Inversion symmetry around the center is present in the staggered model when the number of sites is odd and in the SSH model when the number of sites is even. 

\subsection{Information lattice}

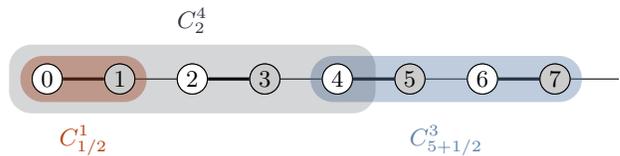
\begin{figure}
    \centering
\definecolor{rust}{HTML}{B7410E}
\definecolor{steelblue}{HTML}{5E81AC}
\definecolor{darkblue}{HTML}{2E3440}
\usetikzlibrary{fit, shapes.geometric}
\begin{tikzpicture}[x=3.375in, y=1cm, 
    siteA/.style={circle, minimum size=4mm, draw=black, fill=white},
    siteB/.style={circle, minimum size=4mm, draw=black, fill=black!20},
    hoppingInter/.style={thin},
    hoppingIntra/.style={-, very thick}
]

% Lattice Parameters
\def\plotWidth{0.9} %Based on matplotlib gs
\def\nCells{3} % Actually # cells - 1
\pgfmathsetmacro{\baseSpacing}{\plotWidth/(\nCells+1)}
\pgfmathsetmacro{\siteSpacing}{\baseSpacing/2}
\def\subLatticeOne{0}
\def\subLatticeTwo{0}
\def\labelpadhop{1.8mm}
\def\labelpadV{1.5mm}
\def\latticeXPad{2mm}
\def\latticeYPad{12mm}
\pgfmathsetmacro{\xOffset}{\plotWidth/10}

\foreach \n in {0,...,\nCells} {
    \pgfmathsetmacro{\xpos}{\xOffset + \n * \baseSpacing}
    \node[siteA] (A\n) at (\xpos - \siteSpacing/2, \subLatticeOne + \latticeYPad) {};
    \node[siteB] (B\n) at (\xpos + \siteSpacing/2, \subLatticeTwo + \latticeYPad) {};
}

%Add invisible node
\pgfmathtruncatemacro{\n}{\nCells+1};
\pgfmathsetmacro{\xpos}{\xOffset + \n * \baseSpacing};
\node[fill=none, draw=none] (A\n) at (\xpos - \siteSpacing/2, \subLatticeOne + \latticeYPad) {};
\draw[hoppingInter] (B\nCells) to node[] {} (A\n);

\foreach \n in {0,...,\nCells} {
    \pgfmathtruncatemacro{\next}{\n+1}
    \ifnum \n<2
        \draw[hoppingIntra] (A\n) to[] node[below=\labelpadhop] {} (B\n);
    \else
        \draw[hoppingIntra] (A\n) to[] node[] {} (B\n);
    \fi 
    \ifnum \n<\nCells
        \ifnum \n<1
            \draw[hoppingInter] (B\n) to node[below=\labelpadhop]  {}  (A\next);
        \else
            \draw[hoppingInter] (B\n) to node[] {} (A\next);
        \fi
    \fi
}

\node[fit=(A0)(B0), draw=none, very thick, fill=rust, rounded corners=3mm, inner xsep=1.5mm, inner ysep=1mm, opacity=0.4] (C1) {};
\node[below=5mm, color=rust] at (C1) {$C_{1/2}^1$};

\node[fit=(A2)(B3), draw=none, very thick, fill=steelblue, rounded corners=3mm, inner xsep=1.5mm, inner ysep=1mm, opacity=0.4] (C2) {};
\node[below=5mm, color=steelblue] at (C2) {$C_{5+1/2}^3$};

\node[fit=(A0)(A2), dashed, draw=none, very thick, fill=darkblue, rounded corners=3mm, inner xsep=3mm, inner ysep=2.5mm, opacity=0.2] (C2) {};
\node[above=5mm, color=darkblue] at (C2) {$C_{2}^4$};

%Redraw the lattice sites to layer them above the subsets
\foreach \n in {0,...,\nCells} {
    \pgfmathsetmacro{\xpos}{\xOffset + \n * \baseSpacing}
    \node[siteA] (A\n) at (\xpos - \siteSpacing/2, \subLatticeOne + \latticeYPad) {};
    \node[siteB] (B\n) at (\xpos + \siteSpacing/2, \subLatticeTwo + \latticeYPad) {};
}

\foreach \n in {0,...,\nCells} {
    \pgfmathtruncatemacro{\even}{2*\n};
    \pgfmathtruncatemacro{\odd}{2*\n+1};
    \node[] at (A\n) {\even};
    \node[] at (B\n) {\odd};
}

\end{tikzpicture}
\caption{Illustration of three subsets $C_n^\ell$ at the left boundary of a one-dimensional lattice. $C^1_{1/2}$ is a strict subset of $C^4_2$, i.e., $C^1_{1/2} \subset C^4_3$, so the information contained in $C^{1}_{1/2}$ is removed when calculating $i^4_3$ according to~\eqref{def:informationlattice}. }
    \label{fig:subsets}
\end{figure}

We give a brief introduction to the information lattice and refer to Refs.~\cite{Kvorning2022,Artiaco_2024,Artiaco25} for details.
The construction of an information lattice starts with a choice of subsystems in which one wants to localize the information in a given quantum state. 
In 1D there is a canonical choice, while in higher dimensions there are multiple possible choices~\cite{Flor2025}.
As in Ref.~\cite{Artiaco_2024}, we denote by $\mathcal{C}^\ell_n$ the set of $\ell+1$ sites centered at $n$. For example, $\mathcal{C}^0_n$ is a single site $n$, while $\mathcal{C}^1_{1/2}$ is the dimer consisting of sites $0$ and $1$ and $\mathcal{C}^{N-1}_{(N-1)/2}$ is the entire system. Figure~\ref{fig:subsets} shows three such subsets near the lattice boundary. The labels $(\ell,n)$ for the subsets $\mathcal{C}^\ell_n$ can be visualized as a triangle, as in Fig.~\ref{fig:small} where every diamond represents a single set $\mathcal{C}^\ell_n$. Note that $n$ is an integer for even $\ell$ and half-integer for odd $\ell$.

The von Neumann information $I^\ell_n$ of a state $\hat{\rho}$ in the subsystem $\mathcal{C}^\ell_n$, which gives the number of bits of information that can be extracted by any set of measurements on $\hat{\rho}$ acting only in the subsystem, is determined by the corresponding reduced density matrix $\hat{\rho}^\ell_n = \Tr_{\bar{\mathcal{C}}^\ell_n} \hat{\rho}$, by
\begin{align}
    I^\ell_n = \ell+1 + \Tr( \hat{\rho}^\ell_n \log_2 \hat{\rho}^\ell_n). \label{eq:Iln}
\end{align}
Here, we use $\log_2$ in order to measure information in bits and we assume a local Hilbert space dimension of $d = 2$ on every site. We use $I$ to denote the total information in the system, i.e., $I=I^{N-1}_{(N-1)/2}$. The information can be seen as the complement of the entropy, which we also measure in bits, $I+S=N$. Both $I$ and $S$ are extensive quantities. 
In a pure state ($S=0$), there is one bit of information per site, while in a mixed state there is less information due to the entropy. 

$I^\ell_n$ describes all the information contained in the sub-system $\mathcal{C}^\ell_n$, even when the information is already there at a smaller scale $\ell'<\ell$. For the information lattice~\cite{Kvorning2022,Artiaco_2024}, we instead work with the information added at scale $\ell$,
\begin{align}
i^\ell_n = I^\ell_n - \sum_{\mathcal{C}^{\ell'}_{n'} \subsetneq \mathcal{C}^\ell_{n}  } i^{\ell'}_{n'}. 
\label{def:informationlattice}
\end{align}
By definition we have the sum rule $I=\sum_{\ell,n} i^\ell_n$, and $i^\ell_n$ describes where in the triangle of Fig.~\ref{fig:small} the information is contained. 
We use $i^\ell=\sum_n i^\ell_n$ to denote a sum over a single row of the triangle, which we interpret as the information contained at scale $\ell$~\cite{Artiaco25} (note that \cite{Artiaco25} denotes this as $I^\ell$ instead of $i^\ell$). 
Figure~\ref{fig:small} shows an example of the information lattice $i^\ell_n$ (colored pyramid) and the corresponding row sums $i^\ell$ for a short SSH chain. 
This construction gives us access to two spatial variables, $\ell$ and $n$. The dependence of $i^\ell$ on the length scale $\ell$ is studied in Sec.~\ref{sec:iell}. The position $n$ within the chain is especially relevant close to the edge of the chain, as will be discussed in Sec.~\ref{sec:ielln}.

As an example, for a single fermionic level with energy $\epsilon$ in the thermal state $\hat{\rho} = n_\text{FD}\ket{1}\bra{1} + (1-n_\text{FD})\ket{0}\bra{0}$, with the Fermi-Dirac occupation function $n_\text{FD} = 1/(e^{\beta(\epsilon - \mu)} + 1)$  and $\beta = 1/(k_BT)$, we have
$S=-n_\text{FD} \log_2(n_\text{FD}) - (1-n_\text{FD}) \log_2(1-n_\text{FD})$ and
$I=1 + n_\text{FD} \log_2(n_\text{FD}) + (1-n_\text{FD}) \log_2(1-n_\text{FD})$. The two terms correspond to the two possible states $\ket{0}$ and $\ket{1}$ and the result is symmetric under the particle-hole transformation $n_\text{FD}\rightarrow 1-n_\text{FD}$. The two pure states, $\hat{\rho} = \ket{0}\bra{0}$ and $\hat{\rho} = \ket{1}\bra{1}$ have zero entropy and full information ($I=1$), while at half-filling $n_\text{FD}=\frac{1}{2}$, $I=0$ and $S=1$. 

For a chain of spinless fermions, $\mathcal{C}^{\ell=0}_n$ corresponds to a single site with a two-dimensional local Hilbert space that has the linearly independent operators $\hat{1}$, $c^\dagger_n$, $c_n$ and $c^\dagger_n c_n$. Assuming $U(1)$ symmetry, the thermal state is a sum of states with a well-defined particle number, so $\langle c^\dagger_n\rangle=\Tr(c^\dagger_n \hat{\rho})=0$ and similarly $\langle c_n\rangle=0$. Thus, the local information $i^{0}_n$ depends on the local electron density $\langle c^\dagger_n c_n\rangle$ only, in the same way as for a single fermionic level, 
\begin{equation}
i^0_n = 1+\langle c^\dagger_n c_n \rangle \log_2(\langle c^\dagger_n c_n \rangle)+(1-\langle c^\dagger_n c_n \rangle)\log_2(1-\langle c^\dagger_n c_n \rangle).\label{eq:ell0}
\end{equation}

\subsection{Information lattice for noninteracting fermions}

The (reduced) density matrix in a spinless fermionic lattice model generally has dimensionality $2^M \times 2^M$ where $M$ is the size of the considered (sub)system. In interacting systems, this makes it challenging to do calculations for large systems and large $M$~\cite{PhysRevLett.122.067203,PhysRevB.110.075115}. However, significant simplifications occur for noninteracting fermions. Detailed derivations are provided by Peschel~\cite{Peschel_2003}, and Cheong and Henley~\cite{cheong2003manybodydensitymatricesfree}. Here, we provide a brief overview aimed at calculating the information lattice, Eq.~\eqref{eq:Iln}, which involves an additional simplification, since we do not need the entire reduced density matrix but only the associated information (entropy), which involves a trace over the spatial indices. The numerical implementation is available at Ref.~\cite{zenodo}.

The main reason for the simplifications is that noninteracting fermionic systems in equilibrium are entirely characterized by the single-particle Green's function $G(i,j)=\langle c_j^\dagger c^{\phantom{\dagger}}_i\rangle$, since all higher-order correlators are then obtained using Wick's theorem. Furthermore, the equilibrium Green's function $G(i,j) = \sum_\alpha n_\alpha G_\alpha(i,j)$ is a sum over the eigenstates $\psi_\alpha(i)$ of the tight-binding Hamiltonian, where $\alpha$ labels the eigenstates, $n_\alpha$ is their (Fermi-Dirac) occupation, and $G_\alpha(i,j)= \psi^\ast_\alpha(j) \psi_\alpha(i)$ their Green's function. 
For the full system, the von Neumann entropy is the sum of the von Neumann entropies of the individual eigenstates, $S=\sum_\alpha S_\alpha=-\sum_\alpha \left[ n_\alpha \log(n_\alpha) + (1-n_\alpha) \log(1-n_\alpha) \right]$. 
Thus, the total entropy does not require access to the eigenfunctions or the spatial structure of the Green's function. 

This is different for the von Neumann entropy $S^A$ of a subsystem $A$ of size $M$, for which we need to restrict the matrix $G(i,j)$ to the subsystem $A$ and calculate its corresponding eigenvalues $g^A_{\zeta}$, which then allows us to evaluate $S^A$ as
\begin{align}
 S^A &= -\sum_{\zeta=1}^M g^A_{\zeta} \log(g^A_{\zeta}) + (1-g^A_{\zeta}) \log(1-g^A_{\zeta}). 
\end{align}
For noninteracting fermions, the chemical potential $\mu$ and temperature $T$ only enter via the Fermi-Dirac occupation numbers $n_\alpha$, whereas the eigenvectors of the full Hamiltonian are independent of these parameters. Since $k_B T\frac{dn_\alpha}{d\mu}=n_\alpha (1-n_\alpha)$, partially filled states are the most responsive parts of a fermionic system; see Appendix~\ref{app:response}. To illustrate this, we plot the function $n_\text{FD} (1-n_\text{FD})$ next to electronic spectra such as those in Fig.~\ref{fig:small}, to give an indication of the relevant energy scale at a given temperature.

\subsection{Metal and insulator}

Since this work concerns the difference between metals and insulators, we should remark on what we mean with this more precisely. For noninteracting electrons at $T=0$ and in the thermodynamic limit, there is a clear distinction between metals and insulators. 
Due to the thermodynamic limit, the bands are continua and the system is a metal when the chemical potential $\mu$ lies in a band and it is an insulator when $\mu$ lies in a band gap. At $T=0$, properties of the system such as the electrical conductivity change discontinuously when $\mu$ crosses the band edge, and this is the metal-insulator transition.

At finite temperature, the thermal broadening of the Fermi-Dirac distribution leads to a continuous crossover instead of a sharp transition between metal and insulator. At sufficiently low $T$, qualitative differences between metal and insulator are visible numerically as long as the chemical potential is not too close to the band edge. At the same time, the concept of a metal is only meaningful when the chain is sufficiently long so that the electronic spectrum appears continuous on the energy scale $k_B T$, so that very long chains are needed to reach the lowest temperatures, eventually putting a limit on what can be simulated accurately. Below, when we talk about metal and insulator, we mean the the finite $T$ and finite size approximation of the true phases at $T=0$ and $N\rightarrow \infty$.

\section{Results}

Figure~\ref{fig:small} shows the information lattice of a short SSH chain at a relatively high temperature. Based on the electronic spectrum, there are six mostly filled levels and two mostly empty ones. 
The $\ell=0$ level shows how much information is contained in local observables, which for the case of spinless fermions considered here is just the electron density, as in Eq.~\eqref{eq:ell0}. 
Most of the information is contained at $\ell=1$ in the dimers with the strong bond $t_1$, as physically expected for the SSH model. 
Due to the open boundary conditions, we see that there is no translation symmetry within a row, but there is an inversion symmetry around the center. For $\ell>1$, there is a rapid decay of $i^\ell$.

\begin{figure}
    \includegraphics{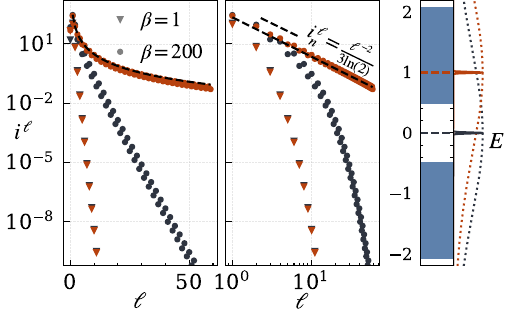}
    \caption{Information per scale $i^\ell$ as a function of the scale $\ell$, shown as a log-plot on the left and as a log-log plot in the middle to identify exponential versus power-law scaling. Here, $N=500$, $t_1=t_2=1$, $V=0.5$. The red data corresponds to $\mu=1$, where the chemical potential lies inside a band and the system is metallic, while the black data is for $\mu=0$, an insulating system with the chemical potential inside the gap. For both chemical potentials, we show high temperature (triangles) and low temperature (circles). The corresponding spectra and the function $n_F (1-n_F)$ are shown for the same high (dotted line) and low (solid line) temperatures. 
    At high temperature (triangles), $i^\ell$ is similar in the the metallic and insulating phase and decays exponentially with $\ell$. At low temperature, the insulator still decays exponentially but with a longer decay length $\xi$, whereas the metal starts to show power-law behaviour. 
    The dashed black line shows the scaling Ansatz.
    Note that the bipartite nature of the system leads to distinct values for even and odd $\ell$.  }
    \label{fig:powerlaw}
\end{figure}

\subsection{Scaling of $i^\ell$}
\label{sec:iell}

Going from the $N=8$ chain of Fig.~\ref{fig:small} to large $N$, the discrete electronic spectrum starts to form bands that become continua in the thermodynamic limit $N\rightarrow \infty$. The position of the chemical potential $\mu$ with respect to the bands determines if the system is a metal or an insulator. Figure~\ref{fig:powerlaw} shows the information per scale $i^\ell$ for two positions of the chemical potential, with the corresponding electronic spectrum shown on the right.  
The information per scale $i^\ell$ decays in all cases with $\ell$, but the decay changes both quantitatively and qualitatively. At high temperatures, the metal and insulator are very similar because the actual occupation numbers of the states are similar, and $i^\ell$ decays exponentially with a short decay length $\xi$. Note that even and odd $\ell$ have a slightly different magnitude due to the dimerized nature of the model. $\xi$ can be extracted from either odd or even sides separately, with numerically similar results. In the insulator, lowering the temperature increases the decay length $\xi$, which can be interpreted as an increased thermal coherence length. On the other hand, in the metal the exponential decay disappears and the decay instead becomes a power law~\cite{Artiaco25}. Specifically, in the metal we expect a scale invariant local information with $i^\ell_n = 1/({3\ln(2)} \ell^{2})$. Since we consider the row-sums in Fig.~\ref{fig:powerlaw}, this corresponds to $i^\ell = ({N-\ell})/({3\ln(2)} \ell^{2})$, which is shown as a dashed line and gives a good match with the numerical results.

The qualitative difference in $i^\ell$ is consistent with how we usually interpret metals and insulators. In a metal at low temperature, and without disorder, electronic quasiparticles can propagate over long distance and this leads to long-ranged phenomena. Viewed differently, the non-analytic step of the Fermi function tends to lead to slower-than-exponential decay in real space. The information per scale is an example of this behavior. 
In contrast, electronic phenomena in insulators are short-ranged~\cite{PhysRev.133.A171} with an energy scale determined by the gap.
At high temperatures, thermal activation in the insulator starts to occur and the difference between the metal and insulator becomes less pronounced. Furthermore, higher temperature leads to shorter correlation lengths.

\begin{figure}
    \includegraphics{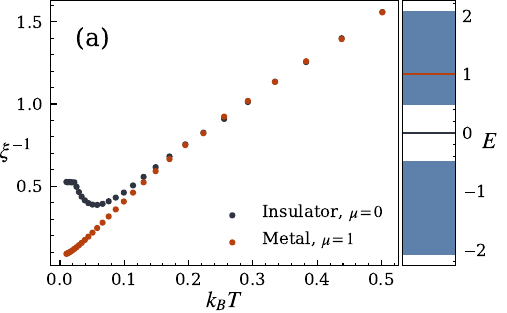}
    \includegraphics[width=0.9\linewidth]{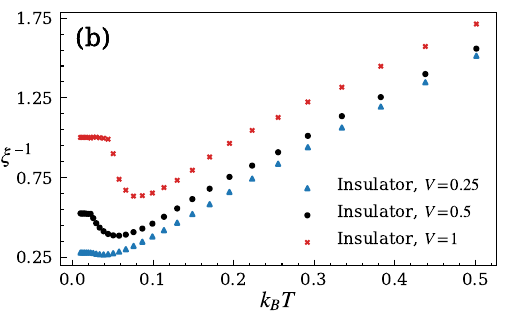}    
    \caption{Inverse decay length $\xi^{-1}$ as a function of $k_B T$, for a chain of length $N=201$ with $t_1=t_2=1$. (a) $V=0.5$ and two different chemical potentials, corresponding to a metal and an insulator as illustrated in the density of states on the right. $\xi$ is obtained by a fitting $i^\ell\sim \exp(-\ell/\xi)$ using even values of $\ell$ only. 
    (b) $V$ controls the size of the gap in the insulator at fixed $\mu=0$, $t_1=t_2=1$ and $N=201$, and therefore has a major impact on the temperature-dependence of $\xi$. 
        }
    \label{fig:xiT}
\end{figure}

Figure~\ref{fig:xiT}(a) shows the correlation length $\xi$ as a function of temperature. 
Coming from high temperature, $\xi$ initially increases as the temperature is lowered (note that $\xi^{-1} $ is plotted), and the metal and insulator have similar $\xi$ at high temperature. As the temperature is lowered, the two phases behave differently. In the metal, $\xi$ keeps increasing in the metal all the way to $T=0$, $\xi(T\rightarrow 0)\rightarrow \infty$. The diverging decay length means that there is no longer exponential decay and that there is power law decay instead (for $\xi \gg N$; for fixed $T$ insulating behavior would eventually return at large enough system size $N \sim \xi$.). 

In the insulator, $\xi^{-1}$ reaches a minimum, then increases and eventually reaches a plateau. Figure~\ref{fig:xiT}(b) shows that the position and height of the plateau are controlled by $V$, which also sets the size of the gap $E_\text{gap}=2V$. Comparing the three cases, the magnitude of $\xi^{-1}$ at the plateau increases with $V$ and the temperature at which the minimum occurs shifts to higher values when the gap is larger. A physical interpretation of this non-monotonous behavior is that for $k_B T\ll E_\text{gap}$, the gap ensures that all the physics is short-ranged, while with increasing temperature it becomes possible to overcome the gap due to thermal fluctuations; at high temperature the decoherence caused by the fluctuations reduces $\xi$ again.

\begin{figure}
\includegraphics[]{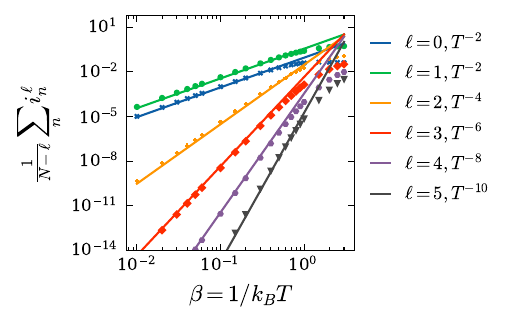}
\caption{Dependence of $i^\ell_n$ on temperature in the SSH model with $t_1=1$, $t_2=1.2$, $\mu=0.75$ and $N=80$. The solid lines show power laws and are drawn as a guide to the eye.}
\label{fig:highT}
\end{figure}

The scaling of the information lattice at high temperature is illustrated in Fig.~\ref{fig:highT}. For each individual scale $\ell$, the numerical results show power-law decay $i^\ell_n \propto T^{-m}$, where $m=2\ell$ for $\ell>0$. The power laws are very robust when the model parameters $t_1$, $t_2$, $V$ and $\mu$ are changed. The case $i^{\ell=0}_n$ is special, since it depends on the density only and it is possible to tune to $i^{\ell=0}_n=0$, independent of temperature.

We do not have a general analytical proof for the $i^\ell_n\sim T^{-2\ell}$ scaling that is observed numerically. However, the particle-hole symmetric trimer ($N=3$) and four-site chain ($N=4$) are small and symmetric enough to solve analytically, while being large enough to $i^{\ell}_n$ for several values of $\ell$. At high temperature, we indeed find $i^{\ell}\sim T^{-2\ell}$, as shown in Appendices~\ref{app:dimertrimer} and \ref{app:foursite}. 

Expressed in terms of $\xi$, the power-law decay of $i^\ell_n$ with temperature gives
\begin{align}
    \exp(1/\xi) &\equiv \frac{i^{\ell}_n}{i^{\ell+1}_{n'}} \propto T^2 \\
    \xi^{-1} &= 2\ln(T)+C \overset{T\rightarrow\infty}{\sim} \ln(T).
\end{align}

For comparison, we note that the 1d Ising model~\cite{baxter2016exactly} has a correlation function at high temperature $\langle \sigma_\ell \sigma_0\rangle = \left[\tanh(J/k_B T)\right]^{\ell}\approx \left(\frac{J}{k_B T}\right)^{\ell}$, which also leads to $\xi^{-1} \sim \ln(T)$. 

\subsection{Cross-sections at fixed $\ell$}
\label{sec:ielln}

In the information lattice $i^\ell_n$, the variable $\ell$ describes the length scale at which information is contained and $n$ the position. In the thermodynamic limit, there is a discrete translation symmetry with period 2, so in the bulk of long chains we expect $i^\ell_n$ to be independent of $n$, apart from the even/odd dichotomy. Since we use open boundary conditions, there will be further $n$-dependence close to the ends of the chain.

\subsubsection{Metal: Friedel oscillations and $k_F$}

\begin{figure}
    \includegraphics{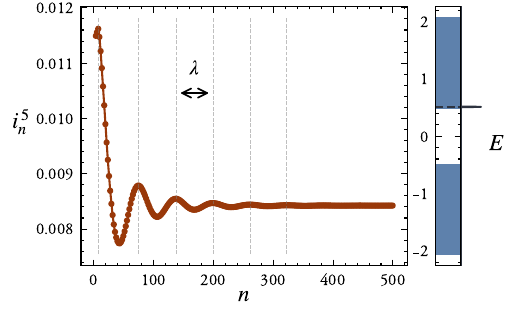}
    \caption{$i^5_n$ as a function of $n$ for the staggered model with $t=1$, $V=0.5$, $\mu=0.51$, $\beta=1000$, and $N=1001$. A clear damped oscillation with wavelength $\lambda$ is seen to emanate from the edge at $n=0$.
    The dashed vertical lines locate the maxima of the oscillation, which can be used to determine the value of $\lambda$.
    Note that only the left half of the system is pictured, i.e., $n\leq 500$, the other half is given by inversion symmetry.
    }
\label{fig:extraction:oscillation}
\end{figure}

\begin{figure}
    \includegraphics{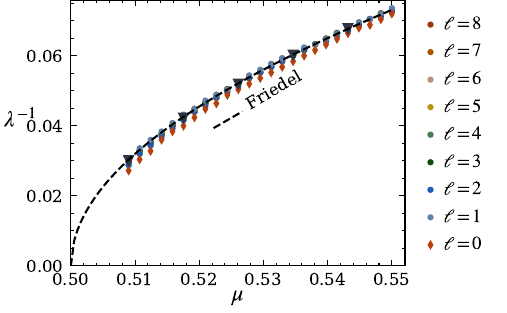}
    \caption{The inverse length scale $\lambda^{-1}$ observed in the information oscillations on a given length scale $\ell$ plotted against $\mu$ for the staggered model with parameters $t=1$, $V=0.5$, $\beta=1000$. A Fourier transform was used to find the dominant frequency in the information oscillations, the inverse of which corresponds to the points depicted. A long chain ($N=5001$) is used for these simulations to enable an accurate determination of $\lambda$ even when it is large. The triangular points correspond to the average distance between maxima in $i^5_n$, as seen in Fig.~\ref{fig:extraction:oscillation}. The dashed line shows $k_F/\pi$, calculated analytically.
    }
    \label{fig:kf}
\end{figure}

The Fermi surface is a central element in our understanding of metals, for example in the form of Landau's theory of the Fermi liquid~\cite{giuliani2008quantum}. Several methods exist to experimentally determine the size and shape of the Fermi surface without needing access to the full electronic spectrum. Some of these are based on oscillations as a function of magnetic field (the de Haas-van Alphen and Shubnikov-de Haas effects), but an alternative is the observation of Friedel oscillations of the electron density around impurities or edges. In the context of the information lattice for tight-binding chains, where $\ell=0$ corresponds to the local density, Friedel oscillations show up in $i^0_n$ and it is natural to ask if they are also present at larger $\ell$. Figure~\ref{fig:extraction:oscillation} indeed shows oscillations in $i^5_n$. The oscillations have a wavelength $\lambda$, as well as overall decay as one gets further from the edge at $n=0$. This is reminiscent of step edges and impurities in STM experiments~\cite{crommie1993imaging}, where this kind of oscillation is used to extract the electronic dispersion. 

Figure~\ref{fig:kf} shows how the wavelength $\lambda$ extracted from these oscillations in $i^\ell_n$ depends on $\mu$. The dashed line shows the expectation for ideal Friedel oscillations, $\lambda=k_F/\pi$, where the analytical dependence of $k_F$ on $\mu$ is given in Appendix~\ref{app:periodic} for the infinite chain at zero temperature. Note that in the one-dimensional chain considered here, the Fermi surface is entirely described by the scalar $k_F$.
We find that $\lambda^{-1}$ is very close to $k_F/\pi$ for all values of $\ell$, although deviations are visible for $\ell=0$ (diamonds). This shows that the information lattice can be used to deduce $k_F$, and that larger values of $\ell$ might even be beneficial for this purpose. 

Regarding the small deviations between the analytical $k_F/\pi$ and the numerical results extracted from $i^\ell_n$, we should note that the formula for Friedel oscillations is derived as the linear response to small perturbations, while the numerical results are based on a rigid end of the chain which is generally beyond the perturbative regime. Furthermore, the analytical results are based on an infinite chain at $T=0$, while the numerical results are for a long but finite chain at low but finite temperature, which can lead to the observed small deviations. 

To understand at what temperature and chemical potential we can expect substantial deviations from the $T=0$ result for Friedel oscillations, we can compare the relevant length scales. 
For the model of a metal considered here, based on physical grounds we anticipate that there are three important length scales: the lattice spacing which is set to 1; the Fermi wavevector $1/k_F$ which depends on $\mu$; and a length scale $\lambda_T$ describing the temperature and the resulting decoherence. The results in Fig.~\ref{fig:kf} are for $\mu$ close to the band minimum $E_0=V=0.5$, so that $1/k_F$ becomes much larger than the lattice spacing, making that length scale irrelevant and leaving us with two length scales. In other words, the lattice fermion model can be replaced by a continuum description in terms of a massive fermion with mass $\frac{1}{m}=\partial^2 E/\partial k^2|_{E_0}$, Fermi wave vector $k_F \approx \sqrt{2m(\mu-E_0)}$ and thermal length scale $\lambda_T=\sqrt{2\pi/mk_B T}$. The competition between temperature-induced decoherence and the Friedel oscillations should be governed by the dimensionless combination $(k_F\lambda_T)^2 = \frac{4\pi (\mu-E_0)}{k_B T}$, with small values corresponding to high temperature and vice versa, and the limit $\mu\rightarrow E_0$ requires special care~\cite{Saha23,Saha25}. However, the smallest values of $\mu$ shown in Fig.~\ref{fig:kf} are $(k_F\lambda_T)^2 \approx 126$ at $\mu=0.51$, which is still quite large, so the Friedel oscillations have not been destroyed by temperature-induced fluctuations. 

\subsubsection{Edge mode in the SSH model}

For the SSH model, there is a clear difference between even and odd chain lengths~\cite{SSHmodel,SSHmodel2,asboth2016short}. The odd chain has an undimerized atom at the end, while the even chain consists fully of dimerized pairs of sites. This is reflected in the electronic spectrum by the presence or absence of a state at $E=0$, in the middle of the bulk gap. In terms of inversion symmetry around the center, the even chain is symmetric and the odd chain is not, which is important for analyzing defect and edge modes~\cite{juan2014bulk-defect-b94,Rhim2017}. 

\begin{figure}
    \centering
    \includegraphics{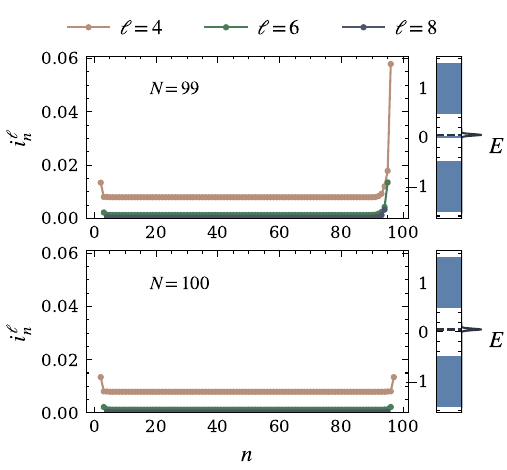}
    \caption{Information distribution of $i_n^\ell$ for three different even length scales $\ell$ in the SSH model with parameters $t_1=1$, $t_2=0.5$, $\mu=0.05$, and $\beta=100$. The first panel shows an odd lattice and the second an even one. The system is an insulator for both cases.
    }
    \label{fig:insulator_ellscale}
\end{figure}

Figure~\ref{fig:insulator_ellscale} shows a comparison of $i^\ell_n$ for an odd and even length SSH chain, with the chemical potential close to the middle of the bulk gap, $\mu=0.05$. 
Whereas $i^\ell_n$ is similar in the bulk for both chains, there are clear differences close to the boundary where $i^\ell_n$ increases. 
The zero-energy-mode in the odd chain is spatially localized close to the edge (at large $n$), and this is reflected in the substantial increase in information there. We observe that this remains visible in the information lattice even at larger values of $\ell$. 
The odd-even chain length dichotomy related to the edge mode is only present when the chemical potential is in the bulk gap, as shown in Appendix~\ref{app:modelcomparison}. 

\section{Conclusion}

In conclusion, we have seen that the metal-insulator transition clearly shows up in the information lattice description, as a change from power-law to exponential scaling of $i^\ell$. At higher temperature, the power law scaling in the metal is suppressed and similar exponential scaling is found for both systems, with faster decay at high temperature. Since $i^\ell$ is an extensive bulk property, we do not find big differences between the staggered and SSH model in the insulating phase.

Edge modes in the information lattice $i^\ell_n$ provide additional insight into these phases. In the metallic phase with small $k_F$, we find oscillations in the information with a wavelength proportional to $k_F^{-1}$, similar to Friedel or RKKY oscillations. These oscillations are present in $i^\ell_n$ at different values of $\ell$, always with the same wavelength. The reason for this is that $k_F^{-1}$ is the only relevant length scale in a noninteracting dilute electron gas. In the insulator, we do not find similar long-wavelength oscillations since there is no Fermi surface. We do see a clear signature of the zero-energy edge mode of odd-length SSH chains in the information lattice.

This study was restricted to metals and band insulators that can be modeled using tight-binding models of noninteracting electrons. The independent nature of the electrons leads to substantial computational simplifications compared to phase transitions driven by interaction~\cite{PhysRevB.109.115104}, disorder~\cite{Artiaco25} or light~\cite{PhysRevB.111.L100302}, which makes it easy to study the information in large systems. 

To study the information lattice for interacting electron models in one dimension, DMRG is the method of choice~\cite{Artiaco25} since it provides access to exact reduced density matrices, up to truncation errors. The higher-dimensional generalization of the information lattice~\cite{Flor2025} could be used to study the Mott metal-insulator in higher dimensions, but this would require numerical methods that give reliable access to the reduced density matrix. Methods based on the imaginary time Green's function are often used for the Hubbard model and few-site reduced density matrices can be extracted from the Green's function and its time-derivatives~\cite{PhysRevB.110.075115}, which enables the calculation of $i^\ell_n$ for small $\ell$, but extending this to $\ell>10$ to study the scaling of the information lattice could be challenging. 

\acknowledgments

EvL and Y.Y. acknowledge support from the Swedish Research Council (Vetenskapsrådet, VR) under grant 2022-03090 and from the Royal Physiographic Society of Lund.
W.S. and E.G. acknowledge the Wallenberg Center for Quantum Technology (WACQT) for financial support via the EDU-WACQT program funded by Marianne and Marcus Wallenberg Foundation. J.H.B. acknolwledges funding from the European Research Council (ERC) under the European Union’s Horizon 2020
research and innovation program (Grant Agreement No. 101001902) and the Knut and Alice Wallenberg Foundation (KAW) via the project Dynamic Quantum Matter (2019.0068). 

\bibliography{refs}

\appendix 

\section{Periodic boundary conditions}
\label{app:periodic}

With periodic boundary conditions, the Hamiltonian of the Rice-Mele tight-binding model is a $2\times 2$ matrix in momentum space,
\begin{align}
H_k = -
\begin{pmatrix}
    -V & t_1 + t_2\exp(ik) \\
    t_1 + t_2\exp(-ik)  & +V
\end{pmatrix}, \label{eq:hmlt:k}
\end{align}
with eigenvalues $\pm \sqrt{V^2+(t_1+t_2 e^{ik})(t_1+t_2 e^{-ik}) }=\pm \sqrt{V^2+t^2_1+t^2_2+2t_1 t_2 \cos(k) } $, i.e., the two bands are symmetric around zero since $\Tr H_k=0$. This dispersion leads to a gap of size $2\sqrt{V^2+(t_1-t_2)^2}$, which simplifies to $2\abs{V}$ for $t_1=t_2$ and to $2\abs{t_2-t_1}$ for $V=0$. Note that the gap opens at $k=\pi$. The other band extrema are at $k=0$ with energy $\pm \sqrt{V^2+(t_1+t_2)^2}$. 

Friedel oscillations and the RKKY interaction have an oscillatory part $\cos(2k_F r+\phi)$ with wavelength $\lambda_\text{Friedel}=\pi/k_F$, i.e., $\lambda_\text{Friedel}^{-1} = k_F/\pi$, which is shown as dashed lines in Figs.~\ref{fig:kf} and \ref{fig:app:wavelength}. Here, $k_F$ is the radius of the Fermi volume. Since the system is studied with the chemical potential just above the band gap and with the band minimum at $k=\pi$, this radius is $k_F =\pi-\arccos\left(\frac{\mu^2-V^2-t_1^2-t_2^2}{2t_1 t_2}\right)$.

A simpler form can be obtained by using the effective mass approximation $E(k)=E_0+\frac{(k-k_0)^2}{2m}$ close to the bottom of the partially filled band. This leads to the standard result $k_F = \sqrt{ 2m (\mu-E_0)}$, which explains the typical square-root behavior seen in the figures. The magnitude of the square root is model-dependent, since $\frac{1}{m}=\frac{\partial^2 E}{\partial k^2}\big|_{k_0}$ for the Hamiltonian in Eq.~\eqref{eq:hmlt:k} leads to
\begin{align}
    \frac{1}{m}\bigg|_{\text{band edge}} &= \frac{t_1 t_2}{\sqrt{V^2+(t_1-t_2)^2}}.
    \label{eq:effectivemass}
\end{align}
For the data in Fig.~\ref{fig:kf}, a staggered model with $t_1=t_2=1$, $V=\frac{1}{2}$, so $2m=1$, which leads to $\lambda_\text{Friedel}^{-1} =\frac{1}{\pi} \sqrt{\mu-E_0}\approx 0.318 \sqrt{\mu-0.5}$. For the SSH model with the same gap, $t_1=1$, $t_2=0.5$, $V=0$,  so $2m=2$ and $\lambda^{-1}_\text{Friedel}\approx 0.450 \sqrt{\mu-0.5}$. 

\section{Information response functions}
\label{app:response}

For a subsystem $A$ of size $M$, we have $S^A=-k_B \sum_{\zeta=1}^M g^A_\zeta \log(g^A_\zeta) + (1-g^A_\zeta) \log(1-g^A_\zeta)$, where $g^A_\zeta$ is an eigenvalue of the matrix $G(i,j)$ restricted to $A$. This directly leads to the derivative
\begin{align}
\frac{dS^A}{dg^A_\zeta}=\log\left(\frac{1-g^A_\zeta}{g^A_\zeta}\right).
\end{align}
Thus, to derive the response of the the von Neumann entropy or the information ($dI^A=-dS^A$) with respect to a physical parameter like $\mu$, one needs to know how the eigenvalues $g^A_\zeta$ change. 

For the noninteracting system considered here, the $M\times M$ Green's function obtained by restricting the full Green's function to the subsystem $A$ corresponds to the thermal Green's function of a fictitious $M\times M$ Hamiltonian $\tilde{H}$, so that $g^A_\zeta=\tilde{n}_\zeta$ is the Fermi-Dirac occupation of the $\zeta$th eigenstate of $\tilde{H}$. Importantly, this fictitious Hamiltonian turns out to be independent of $T$ and $\mu$, so that we can use the derivative of the Fermi-Dirac distribution,
\begin{align}
    \frac{dg^A_\zeta}{d\mu}=\frac{d\tilde{n}_\zeta}{d\mu} = \beta \tilde{n}_\zeta (1-\tilde{n}_\zeta)= \beta g^A_\zeta (1-g^A_\zeta).
\end{align}
Note that this only works for noninteracting fermions, for interacting fermions it is possible to find a description of a restricted Green's function at constant value of the parameters in terms of an auxiliary Hamiltonian, but the auxiliary Hamiltonian will change when $\mu$ changes and one therefore gets an additional term $d\tilde{H}/d\mu$, i.e., a two-particle correlation function.

In total, the response of the information to changes in the chemical potential is
\begin{align}
        \frac{d}{d\mu}I_n^\ell 
        &=\beta\sum_\zeta (1-g^{\ell,n}_\zeta)g^{\ell,n}_\zeta \log\left(\frac{g^{\ell,n}_\zeta}{1-g^{\ell,n}_\zeta}\right).
\end{align}

\section{Analytical solution for a trimer}
\label{app:dimertrimer}

To get more intuition for the information lattice for fermionic systems, it is instructive to analytically solve the trimer with $t_1=t_2=t$, $V=0$ and $\mu=0$. This choice of parameters ensure particle-hole symmetry and therefore leads to relatively simple results. We have $H=-t\begin{pmatrix} 0&1&0\\1&0&1\\0&1&0 \end{pmatrix}$ with eigenvalues $\lambda_\pm = \pm t \sqrt{2}$ and $\lambda_0=0$, with eigenvectors $\frac{1}{2} (1, -\sqrt{2},1)$, $\frac{1}{2} (1, \sqrt{2},1)$ and $\frac{1}{\sqrt{2}} (1,0,-1)$. Denote $n_\text{FD}(\lambda_+)=x$, so that $n_\text{FD}(\lambda_-)=1-x$ by particle-hole symmetry, and $n_\text{FD}(0)=\frac{1}{2}$ independent of temperature. As a result, the Green's function is
\begin{align}
    G =& \frac{x}{4} \begin{pmatrix} 1 & -\sqrt{2} & 1 \\ -\sqrt{2} & 2 & -\sqrt{2} \\ 1 & -\sqrt{2} & 1 \end{pmatrix} + \frac{1-x}{4} \begin{pmatrix} 1 & \sqrt{2} & 1 \\ \sqrt{2} & 2 & \sqrt{2} \\ 1 & \sqrt{2} & 1 \end{pmatrix} \notag \\ &+\frac{1}{4} \begin{pmatrix} 1 & 0 & -1 \\ 0 & 0 & 0 \\ -1 & 0 & 1 \end{pmatrix} \\
    =& \frac{1}{2}\hat{1} + \frac{\sqrt{2}(1-2x)}{4} \begin{pmatrix} 0 & 1 & 0 \\ 1 & 0 & 1 \\ 0 & 1 &0 \end{pmatrix}, \label{eq:app:trimer:gf}
\end{align}
which has eigenvalues $\frac{1}{2}$, $1-x$ and $x$ as expected, since these are the occupation numbers of the eigenstates. The first eigenvalue contributes zero to the total information, the two other eigenvalues give an identical contribution proportional to $(x-\frac{1}{2})^2\sim T^{-2}$ at high temperature, where $x$ is close to $\frac{1}{2}$. More precisely,
\begin{align}
    I=I^{\ell=2}_n = 2+2x \log_2(x)+2(1-x) \log_2(1-x).
\end{align}

For $I^{\ell=0}_n$, the $1\times 1$ subblocks on the diagonal of Eq.~\eqref{eq:app:trimer:gf} are used, which are all equal to $\frac{1}{2}$ and therefore contribute zero to the information (half-filling), so $I^{\ell=0}_n=0$. 

For $I^{\ell=1}_n$, the upper and lower $2\times 2$ block of Eq.~\eqref{eq:app:trimer:gf} are used, which are identical by symmetry. This block $\begin{pmatrix} \frac{1}{2} & \frac{\sqrt{2}}{4}(1-2x) \\ \frac{\sqrt{2}}{4}(1-2x) & \frac{1}{2}\end{pmatrix}$ has eigenvalues $\frac{1}{2} \pm \frac{(x-\frac{1}{2})\sqrt{2}}{2}$, so 
\begin{align}
    I^{\ell=1}_n= 2 &+ 2\left(\frac{1}{2} + \frac{(x-\frac{1}{2})}{\sqrt{2}}\right) \log_2\left(\frac{1}{2} + \frac{(x-\frac{1}{2})}{\sqrt{2}}\right) \notag \\
    &+ 2\left(\frac{1}{2} - \frac{(x-\frac{1}{2})}{\sqrt{2}}\right) \log_2\left(\frac{1}{2} - \frac{(x-\frac{1}{2})}{\sqrt{2}}\right).
\end{align}
From these results, it follows that the information lattice $i^\ell_n$ is given by 
\begin{align}
    i^{\ell=0}_n&=0 \\
    i^{\ell=1}_n&= I^{\ell=1}_n \\
    i^{\ell=2}_n&= I^{\ell=2}_n-2 I^{\ell=1}_{n'}+I^{\ell=0}_n, 
\end{align}
where there is no dependence on $n$ in this small, highly symmetric set-up. The formulas for $I^{\ell=2}_n$ and $i^{\ell=2}_n$ simplify in the high-temperature limit where $x=\frac{1}{2}+\delta x$, with $\delta x\approx - \frac{t}{2\sqrt{2} k_BT}$ so that.
\begin{align}
    I^{\ell=2}_n &\approx \frac{4}{\ln(2)} \left(\delta x\right)^2+\frac{8}{\ln(8)} \left(\delta x\right)^4 + \ldots \\
    I^{\ell=1}_n &\approx \frac{4}{\ln(2)} \left(\frac{\delta x}{\sqrt{2}}\right)^2+\frac{8}{\ln(8)} \left(\frac{\delta x}{\sqrt{2}}\right)^4 + \ldots \\ 
    i^{\ell=1}_n &\approx \frac{2}{\ln(2)} (\delta x)^2= \frac{2}{\ln(2)} \left(\frac{\sqrt{2} t}{4k_B T}\right)^2 = \frac{t^2}{4 \ln(2) k_B^2 T^2}\notag  \\
    i^{\ell=2}_n &\approx \frac{4}{\ln(8)} (\delta x)^4= \frac{4}{\ln(8)} \left(\frac{\sqrt{2} t}{4k_B T} \right)^4 = \frac{t^4}{16 \ln(8) k_B^4 T^4} \notag \\
\end{align}

Altogether, for the particle-hole symmetric trimer, the leading term in the total information at high temperature is $I\sim T^{-2}$ and this leading term is entirely contained at the $\ell=1$ level (no information at $\ell=0$ by symmetry). The $\ell=2$ layer of the information lattice only contributes with a sub-leading $T^{-4}$ term. If we define $\xi$ as
\begin{align}
    \exp(1/\xi) &\equiv \frac{i^{\ell=1}_n}{i^{\ell=2}_{n'}} = \frac{12 k_B^2 T^2}{t^2}    
    \\
    \xi &= \frac{1}{\ln(12 k_B^2 T^2/t^2)} \overset{T\rightarrow\infty}{\sim} \mathcal{O}\left(\frac{1}{\ln(T)}\right),
\end{align}
then we find a logarithmic behavior of $\xi$ at high temperature.

\section{Four-site chain}
\label{app:foursite}

The derivation of the information lattice for the four-site chain proceeds along similar lines as the trimer in Appendix~\ref{app:dimertrimer}. We consider $t_1=t_2=t$, $V=0$ and $\mu=0$ for simplicity, 
        $E_1 = \frac{-1- \sqrt{5}}{2}t$, $E_2 = \frac{1-\sqrt{5}}{2}t$, $E_3 = \frac{-1+\sqrt{5}}{2}t$ $E_4 = \frac{1+\sqrt{5}}{2}t$,
and corresponding eigenvectors
\begin{align}
    \begin{split}
        \ket{\Psi_1}&=\frac{1}{\sqrt{5+\sqrt{5}}}\left(1,\frac{1}{2}(1+\sqrt{5}),\frac{1}{2}(1+\sqrt{5}),1\right),\\
        \ket{\Psi_2}&=\frac{1}{\sqrt{5-\sqrt{5}}}\left(-1, \frac{1}{2}(1 - \sqrt{5}), -\frac{1}{2} (1 - \sqrt{5}), 1\right),\\
        \ket{\Psi_3}&=\frac{1}{\sqrt{5-\sqrt{5}}}\left(1 , \frac{1}{2} (1 - \sqrt{5}), \frac{1}{2}(1 - \sqrt{5}), 1\right),\\
        \ket{\Psi_4}&=\frac{1}{\sqrt{5+\sqrt{5}}}\left(-1, \frac{1}{2} (1 + \sqrt{5}), -\frac{1}{2}(1 + \sqrt{5}), 1\right).
    \end{split}
\end{align}
By particle-hole symmetry,
\begin{align}
    n_{FD}(E_1)&=1-n_{FD}(E_4)\equiv x, \\ 
    n_{FD}(E_2)&=1-n_{FD}(E_3) \equiv y,
\end{align}
and the Green's function after simplification is
\begin{align}
    \begin{split}
        G&=\frac{x}{5+\sqrt{5}}\left(
\begin{array}{cccc}
 0 & 1+\sqrt{5} & 0 & 2 \\
 1+\sqrt{5} & 0 & 3+\sqrt{5} & 0 \\
 0 & 3+\sqrt{5} & 0 & 1+\sqrt{5} \\
 2 & 0 & 1+\sqrt{5} & 0 \\
\end{array}
\right)\\
&+\frac{y}{5-\sqrt{5}}\left(
\begin{array}{cccc}
 0 & \sqrt{5}-1 & 0 & -2 \\
 \sqrt{5}-1 & 0 & \sqrt{5}-3 & 0 \\
 0 & \sqrt{5}-3 & 0 & \sqrt{5}-1 \\
 -2 & 0 & \sqrt{5}-1 & 0 \\
\end{array}
\right)\\
&+\frac{1}{2\sqrt{5}}\left(
\begin{array}{cccc}
 \sqrt{5} & -2 & 0 & 1 \\
 -2 & \sqrt{5} & -1 & 0 \\
 0 & -1 & \sqrt{5} & -2 \\
 1 & 0 & -2 & \sqrt{5} \\
\end{array}
\right).
    \end{split}
\end{align}
with the expected eigenvalues $1-x$, $x$ ,$1-y$, and $y$ representing the occupancy of the eigenstates.

For $\ell=0$, the eigenvalues of the $1\times 1$ blocks are the diagonal elements of the Green's function, which are all $\frac{1}{2}$ and hence do not contain information due to half filling, so $I_n^{\ell=0}=0$.

For the length scale $\ell=3$ (the entire chain),  the information is
\begin{align}
    \begin{split}
        I&=I_{1+1/2}^{\ell=3}\\&=4+ 2[x\log_2 x+(1-x)\log_2(1-x)]  \\ &\phantom{=4}+2[y\log_2 y+(1-y)\log_2(1-y)].
    \end{split}
\end{align}
For $\ell=2$, the eigenvalues of the Green's function are
\begin{align}
    \lambda_1 = \frac{1}{2}, \,\, \lambda_{2,3} = \frac{1}{2}\pm \alpha(x,y), 
\end{align}
where
\begin{align}
    \alpha&=\frac{\sqrt{5}}{10}\sqrt{2 \left(5+\sqrt{5}\right) (x-1) x-2 \left(\sqrt{5}-5\right) (y-1) y+5}.
\end{align}
Then, the information is given by
\begin{align}
    \begin{split}
        I_n^{\ell=2}=2&+2\left(\frac{1}{2}-\alpha(x,y)\right)\log \left(\frac{1}{2}-\alpha(x,y)\right) \\ 
        &+2\left(\frac{1}{2}+\alpha(x,y)\right)\log \left(\frac{1}{2}+\alpha(x,y)\right).
    \end{split}
\end{align}
For $\ell=1$ (2 sites), there are two distinct cases to consider, $I^{\ell=1}_{n=1/2}$ and $I^{\ell=1}_{n=1+1/2}$, corresponding to the edge and the center, respectively. 
For the edge, the Green's function matrix has eigenvalues
\begin{align}
    \lambda_{1,2} =\frac{1}{2}\pm \gamma(x,y) \spacecomma \gamma(x,y)=\frac{\sqrt{(x+y-1)^2}}{\sqrt{5}},
\end{align}
resulting in information
\begin{align}
    \begin{split}
        I_{n=1/2}^{\ell=1}&=2+2\left(\frac{1}{2}+\gamma(x,y)\right)\log_2 \left(\frac{1}{2}+\gamma(x,y)\right)\notag \\ &\phantom{=2+}+2\left(\frac{1}{2}-\gamma(x,y)\right)\log_2 \left(\frac{1}{2}-\gamma(x,y)\right)
    \end{split}
\end{align}
For the central block, the eigenvalues are
\begin{align}
   \lambda&=\frac{1}{2}\pm \kappa(x,y), \\ \kappa(x,y) &= \frac{1}{10}\left(\sqrt{5}-\left(5+\sqrt{5}\right) x+\left(\sqrt{5}-5\right) y\right),
\end{align}
resulting in information
\begin{align}
    \begin{split}
        I_{n=1+1/2}^{\ell=1}&=2+2\left(\frac{1}{2}+\kappa(x,y)\right)\log_2 \left(\frac{1}{2}+\kappa(x,y)\right)  \\ &\phantom{2+}+2\left(\frac{1}{2}-\kappa(x,y)\right)\log_2 \left(\frac{1}{2}-\kappa(x,y)\right).
    \end{split}
\end{align}
Hence, the information added at each scale is
\begin{align}
    \begin{split}
        i_n^{\ell=0}&=0,\\
        i^{\ell=1}_{n}&=I_n^{\ell=1},\\
        i_n^{\ell=2}&=I^{\ell=2}_n-I_{n-1/2}^{\ell=1}-I_{n+1/2}^{\ell=1},\\
        i^{\ell=3}_n&=I_n^{\ell=3}-2I^{\ell=2}_{n=1}+I_{n=1+1/2}^{\ell=1}.
    \end{split}
\end{align}
where in the last line we used $I^{\ell=1}_{n=1/2}=I^{\ell=1}_{n=2+1/2}$. 

In the high temperature limit, the Fermi-Dirac distribution is expanded as
\begin{align*}
    x &= n_{FD}(E_1)=\frac{1}{2}+\delta x  \spacecomma \delta x\approx -\frac{E_1}{4k_BT}=\frac{\sqrt{5}+1}{8k_BT}t\\
    y &= n_{FD}(E_2)=\frac{1}{2}+\delta y \spacecomma \delta y \approx -\frac{E_2}{4k_BT}=\frac{\sqrt{5}-1}{8k_BT}t.
\end{align*}
Defining
$        \delta \kappa = \kappa\left(\frac{1}{2}+\delta x,\frac{1}{2}+\delta y\right) $
and similarly $\delta \gamma$, $\delta \alpha$,
\begin{align}
   \begin{split}
      \delta \kappa&=-\frac{t}{4k_B T} \spacecomma \delta \gamma=\frac{t}{4k_B T} \spacecomma \delta \alpha =- \frac{t}{2\sqrt{2}k_B T}
   \end{split}.
   \notag
\end{align}
The Taylor expansion of the information has general form
\begin{align}
    I_{\text{general}}(z)&=C \notag \\
    &+2\left(\frac{1}{2}+z\right)\log_2 \left(\frac{1}{2}+z\right) \notag \\ 
    &+2\left(\frac{1}{2}-z\right)\log_2 \left(\frac{1}{2}-z\right)\\
    &\approx  C-2+\frac{4z^2}{\ln(2)}+\frac{8z^4}{\ln(8)}+\frac{64 z^6}{15\ln(2)},
\end{align}
which only contains even powers of $z$.
Then,
\begin{align}
    \begin{split}
        I_n^{\ell=3}&=\frac{4}{\ln(2)}[(\delta x)^2+(\delta y)^2]+\frac{8}{\ln(8)}[(\delta x)^4+(\delta y)^4]+\dots\\
        I_n^{\ell=2}&= \frac{4}{\ln(2)}(\delta \alpha)^2+\frac{8}{\ln(8)}(\delta \alpha)^4+\dots\\
        I_{n}^{\ell=1}&=\frac{4}{\ln(2)}(\delta \gamma)^2+\frac{8}{\ln(8)}(\delta \gamma)^4+\dots,
    \end{split}
\end{align}
since $(\delta \gamma)^2=(\delta \kappa)^2$. The information added at each scale is then
\begin{align}
    i^{\ell=1}_{n}& \approx \frac{t^2}{4\ln(2)k_B^2T^2},\\
    i_{n}^{\ell=2}&\approx \frac{t^4}{48\ln(2) k_B^4T^4},\\
    i_n^{\ell=3}&\approx \frac{t^6}{320 \ln(2) k_B^6T^6}.
\end{align}
As for the trimer, $i^\ell \sim T^{-2\ell}$.

\section{Finite size effects}
\label{app:finitesize}

\begin{figure}
    \includegraphics{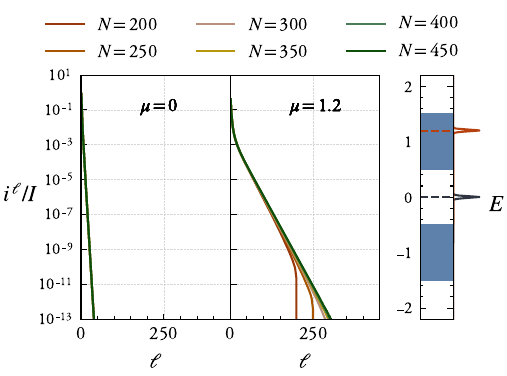}    
    \caption{Role of finite-size effects, in the SSH Hamiltonian with $t_1=1$, $t_2=0.5$, $V=0$,  $\beta=100$. In the insulator (left), finite-size effects are generally small until $\ell \approx N$. In the metal, finite-size effects set in earlier and large system sizes are needed for the simulations. The band structure for $N=200$ is displayed in the right panel. $i^\ell$ shown on odd rows only to avoid staggering. The information per row is normalized through division by the total information in the system. 
    }
    \label{fig:finitesize}
\end{figure}

Given the slow decay of $i^\ell$, especially in the metal at low temperature, it is important to study the finite-size convergence of the simulations. Note that the information per row, $i^\ell$, is an extensive quantity so that normalization is needed, here we normalize by the total information $I=\sum_\ell i^\ell$ which is also extensive. Figure~\ref{fig:finitesize} shows that the results in the insulating phase are generally weakly dependent on the size of the system, due to the overall fast spatial decay in insulating systems. The situation is more subtle in the metal at low temperature, where the information decays slowly with $\ell$. By construction, we  need $\ell < N$, so finite size effects become very visible for $\ell\approx N$. At this temperature, only minor finite size effects are visible for $\ell \ll N$, showing that the results at this temperature are converged. Generally, the approximation of a continuous spectrum by discrete peaks leads to finite size effects when the temperature is small compared to the discretizations of the energies. 

\begin{figure}
    \centering
    \includegraphics{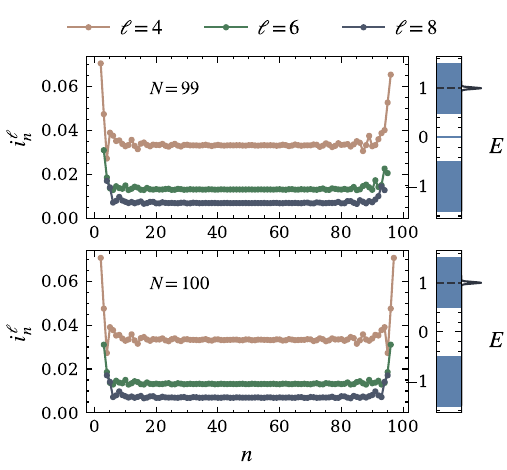}
    \caption{Information distribution of $i_n^\ell$ for three different even length scales $\ell$ in the SSH model with parameters $t_1=1$, $t_2=0.5$, $\mu=1$, and $\beta=100$. The first panel shows an odd lattice and the second an even one. The system is metallic for both cases.}
    \label{fig:metal_ellscale}
\end{figure}

\section{Comparison of SSH and staggered model}
\label{app:modelcomparison}

\begin{figure}
    \centering
    \includegraphics{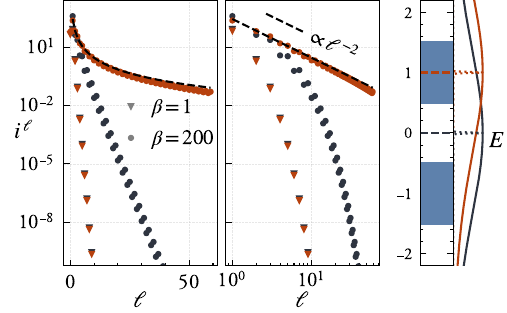}
    \caption{Decay in the metal and insulator, for the SSH model, similar to Fig.~\ref{fig:powerlaw} for the staggered model. The parameters in both models are set so that the gap is equal, which means that the SSH model has a smaller band width compared to the staggered model. In both cases, $\ell^{-2}$  power law decay is found in the metal at low temperature, versus exponential decay in all other cases, but the decay lengths are different. }
    \label{fig:app:powerlaw}
\end{figure}

\begin{figure}
    \includegraphics{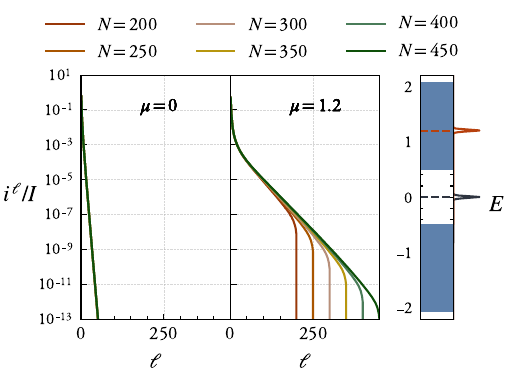}
    \caption{Finite size effects, staggered potential $t_1=t_2=1,V=0.5,\beta=100$.  The information is normalized through division by the total information in the system. The corresponding SSH model is shown in Fig.~\ref{fig:finitesize}.}
\label{fig:finite:staggered}
\end{figure}

\begin{figure}
    \centering
    \includegraphics{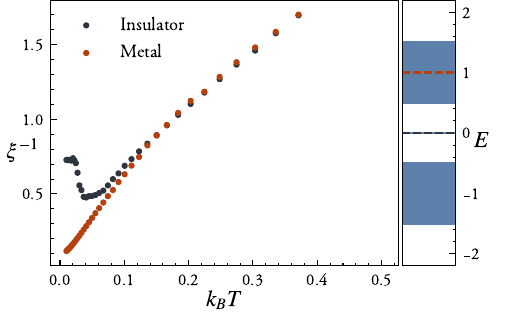}
    \caption{
    Decay length versus temperature in the SSH model for even $N$ and with parameters $t=1$, $V=0.5$, which gives the same gap as in Fig.~\ref{fig:xiT}.}
    \label{fig:app:xiT2}    
\end{figure}

\begin{figure}
    \includegraphics{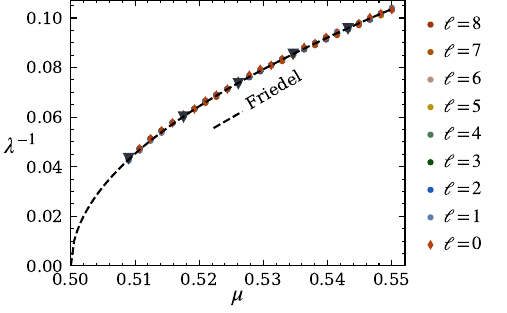}
    \caption{The inverse length scale $\lambda^{-1}$ observed in the information oscillations on a given length scale $\ell$ plotted against $\mu$ for the SSH model with parameters $t_1=1$, $t_2=0.5$, $\beta=1000$, $N=5000$. A Fourier transform was used to find the dominant frequency in the information oscillations, the inverse of which corresponds to the points depicted. The dashed line shows the exact $k_F/\pi$.}
    \label{fig:app:wavelength}
\end{figure}

Both the staggered on-site potential $V$ in the staggered model and the staggered hopping parameters $t_1$, $t_2$ in the SSH model create a gap, but the mechanism behind the gap is different. Thus, there can be differences in the information lattice related to the different character of the insulating phase. To make a standardized comparison, we fix the magnitude of the gap ($2\abs{V}$ in the staggered model and $2\abs{t_1-t_2}$ in the SSH model) and the strong hopping parameter ($t_1=1$ in the SSH model, $t=t_1=t_2=1$ in the staggered model). This does mean that the bandwidth and effective mass of the models are different.

The decay of $i^\ell$ at fixed temperature in the SSH model is shown in Fig.~\ref{fig:app:powerlaw}, cf. \ref{fig:powerlaw} in the main text. Finite-size effects in the staggered model are shown in Fig.~\ref{fig:finite:staggered}, cf. Fig.~\ref{fig:finitesize}.
The temperature-dependence of $\xi$ for models with the same gap is shown in Fig.~\ref{fig:xiT}(a) and Fig.~\ref{fig:app:xiT2} for the staggered and SSH model, respectively. 

Overall, these results show that the two models behave qualitatively similarly for comparable gaps, but quantitative differences are visible. For example, in the insulator at low temperature in Figs.~\ref{fig:xiT}(a)-\ref{fig:app:xiT2}, $\xi^{-1}$ in the staggered model has substantially lower values overall. A perfect agreement cannot be expected since the models have different electronic wavefunctions and spectra. In particular, the effective mass is different. 

The metallic oscillations with scale $\lambda\sim k_F^{-1}$ are also present in the metallic phase of the SSH model, as shown in Fig.~\ref{fig:app:wavelength}. Due to the difference in effective mass, the prefactor of the scaling $\lambda^{-1}\sim \sqrt{\mu-E_0}$ is different in the SSH model, as discussed in Appendix~\ref{app:periodic}. Unlike for the insulating phase of the SSH model, we do not find a strong difference between a chain of even and odd length in the metallic phase, see Fig.~\ref{fig:metal_ellscale}. In terms of the electronic structure, the difference between the even and odd length chain is the presence of a protected mode at $E=0$. Since the chemical potential is very far away from $E=0$ in this case, it is not unexpected that the existence of this mode does not substantially change the information lattice.

Note that a quantitative difference in the $\ell=0$ mode is clearly visible when comparing Figs.~\ref{fig:kf} and \ref{fig:app:wavelength}. In the SSH model, the electron density per site is constant in the bulk of the chain and only varies close to the edge. In the staggered model, it has oscillations with wavelength 2 even in the bulk, due to the explicitly changing on-site energy. These oscillations are reflected in $i^{0}_n$ and make $\ell=0$ rather special in the staggered model.

\end{document}